\documentclass[11pt,leqno]{article}   

\usepackage[T1]{fontenc}
\usepackage[utf8]{inputenc}
\usepackage{authblk}
\usepackage[all]{xy}
\usepackage{authblk}
\usepackage[symbol*]{footmisc}

\usepackage{amsmath,amstext,amsthm,amsfonts,amssymb,color,latexsym}
\numberwithin{equation}{section}

\def\N{\mathbb{ N}}

\newcommand{\ic}{\text{i}} 

\usepackage{epsfig}

\DefineFNsymbolsTM{otherfnsymbols}{%
  \textbullet \circ
  \textsection   \mathsection
  \textdagger    \dagger
  \textdaggerdbl \ddagger
  \textasteriskcentered *
  \textbardbl    \|%
  \textparagraph \mathparagraph
}%

\setfnsymbol{otherfnsymbols}

\def\N{\ensuremath{\mathbb N}}

\newcommand{\be}{\begin{equation}}
\newcommand{\ee}{\end{equation}}
\newcommand{\bea}{\begin{eqnarray}}
\newcommand{\eea}{\end{eqnarray}}
\newcommand{\beano}{\begin{eqnarray*}}
\newcommand{\enano}{\end{eqnarray*}}

\newcommand{\ena}{\end{eqnarray}}

\newcommand{\en}{\end{equation}}

\newcommand{\ba}{\begin{array}}
\newcommand{\ea}{\end{array}}

\newcommand{\bg}{\begin{gathered}}
\newcommand{\eg}{\end{gathered}}

\title{A Note on Deformed Ladder Operators for Noncommutative Morse Oscillator}
\author[a]{Nadhira A. H. \thanks{nharis27@gmail.com}}
\author[a,b]{Nurisya M.S.
\thanks{risya@upm.edu.my}}
\author[a,b]{K.T. Chan}

\affil[a]{Laboratory of Computational Sciences \& Mathematical Physics, Institute for Mathematical Research, Universiti Putra Malaysia, 43400 UPM Serdang, Selangor, Malaysia}
\affil[b]{Department of Physics, Faculty of Science, Universiti Putra Malaysia, 43400 UPM Serdang, Selangor, Malaysia}

\providecommand{\keywords}[1]{\textbf{\textit{Keywords:}} #1}

\begin{document}

\maketitle
\begin{abstract}
Morse oscillator is one of the known solvable potentials which attracts many applications in quantum mechanics especially in quantum chemistry. One of the interesting results of this study is the generation of ladder operators for Morse potential. The operators are a representation of the shifting energy levels of the states exhibited by the wavefunction. From this result, we manipulate and deform the operators in such a way that it gives a noncommutative property to promote noncommutative quantum mechanics (NCQM). The resultant NC feature can be shown in the spatial coordinates and finally the Hamiltonian. In this study, we consider two-dimensional Morse potential where the ladder operators are in the form of the corresponding 2D Morse. \\


\end{abstract}


\section{Introduction}

Morse potential has been in the spotlight for it is a more realistic potential to describe diatomic molecular behaviour as opposed to the classic harmonic oscillator. Many are interested in this model to work on the molecular spectroscopy better. Therefore, most studies on Morse oscillator can be found in the discipline of quantum chemistry where the concern is mainly focused on the physical potential behaviour of molecules. One example is shown in~\cite{mccoy}. Previous research including~\cite{shi} and~\cite{avram} are in the same direction as our study, which is getting the grips with the problem through the algebraic approach. We realised that it is somehow interesting to associate the noncommutative quantum mechanics (NCQM) with this since the existing methods to solve for NCQM oftentimes involve the use of Moyal products and supersymmetric quantum mechanics (SUSYQM). We propose our work tackling the problem from the perspective of ladder operators which later can be directed to the whole representation of the system. Our take on Morse oscillator is primarily to review the potential and deform it as a means of conducting NCQM. 

Noncommutative quantum mechanics (NCQM) has been widely studied and for one of many reasons, its importance is seen to be related with quantum potentials. Quite a number of one-dimensional problems have been explored and they are proven to be solvable. Some work are as featured in~\cite{shi},~\cite{avram} and~\cite{bordoni}. Therefore, many are attempting to further the study to higher dimensional cases and the different ways of solving them. In quantum potentials, the 1D cases are usually analytically solvable, while 2D cases can be solved by separation of variables. For most solutions of 2D and 3D problems, the Schr$\ddot{o}$dinger equation is workable by using reduction method and oftentimes, irreducible to 1D problems are tough to solve~\cite{tichy}. A study of a generalised case of 3D Morse potential has been done by~\cite{avram} that includes anharmonicity of the potential. It is worth to note that Morse potential possesses the anharmonicity as part of its system and as highlighted in~\cite{avram}, weak anharmonicity can be detected in 3D Morse oscillator when there is only one direction that the oscillation is anharmonic while other two directions harmonic. 

Findings due to NCQM research are mostly involved with the case of stationary quantum states together with its associated the so-called deformed or generalised energy levels parameterised with NC parameters. For this case, solutions arise in the NC structure are constructed from solving the time-independent Schr$\ddot{o}$dinger equation. For the past decades, at least to the authors' knowledge, there is plenty of literature analysing the nontrivial case of examples of NCQM models with regard to the stationary states. Many successful cases highlighted include the Landau level problems, see for example~\cite{isiaka} and the references therein. A good review on the formulation aspects of establishing NCQM model via different transformation methods is properly discussed in~\cite{gouba}.
For the case of non-stationary states due to NCQM, a review on the dynamical aspects of noncommutative system can be obtained from~\cite{bemfica} while other work reported involves the energy-dependent Schrodinger equation~\cite{herko}. 
Another recent interest on NCQM is~ \cite{bertolani} for which one of the ideas is the importance of NCQM deformations in further understanding the transition of quantum to classical interpretation and exploiting non-Hermition NCQM using Dyson and Seiberg-Witten maps to utilise in canonical quantum mechanics ~\cite{santos}.

To grasp the idea of NCQM, one has to be aware of how it distinguishes from the standard quantum mechanics. The apparent element to differentiate noncommutative quantum mechanics from the standard quantum mechanics is the association of additional parameters to the commutation relations for position and momentum operators. The ordinary canonical commutation relations for operators 
\be
[x_{i},x_{j}] =0, \ \  \quad [p_{i},p_{j}] =0, \ \  \quad [x_{i},p_{j}]= \ic\hbar\delta_{ij},
\ee
are deformed such that
\be\label{e:ncom}
[\hat{x}_{i},\hat{x}_{j}] =  \ic\theta_{ij}, \quad [\hat{p}_{i},\hat{p}_{j}] =\ic\zeta_{ij}, \quad [\hat{x}_{i},\hat{p}_{j}]= \ic\hbar\delta_{ij}.
\ee
where $\theta_{ij}$ and $\zeta_{ij}$ are anti-symmetric real values of dimension of (length)$^2$. Here, the NC operators are defined in the same Hilbert space as the commutative ones \cite{Bertolami}. Most models of NCQM are the extension of canonical Heisenberg algebra \cite{IN11} and the importance of maintaining the canonical Heisenberg commutation relation for NCQM model is so that it is consistent with ordinary commutative quantum mechanics \cite{IN10}. The idea of translating the mathematical means of classical mechanics to quantum mechanics is realisable through the transformation of the commutative algebra of classical observables to that of quantum mechanical observables which are noncommutative. The obvious difference of these two is that in order to show the noncommutative property of the algebra, it has to be at least in two dimensions. (\ref{e:ncom}) shows the noncommutative algebra of both spatial and momentum operators with $\theta_{ij}$ and $\zeta_{ij}$ being the non-zero noncommutative parameters while $\delta_{ij}$ is the standard Kronecker delta for $i,j=1,2$. One may suggest that (\ref{e:ncom}) serves as translation operators \cite{IN1} that shift particles and fields by a certain magnitude in a certain direction. Interestingly, the purpose of algebraic or commutation relations is that they implement uncertainty relations \cite{duplicher}.  

In this work, we show the construction of the noncommutative Morse oscillator for which the idea is to study the mathematical structure of noncommutativity features in terms of a diatomic quantum potential. We propose another way (deformed operator) of transforming the usual Morse ladder operator instead of the common method that tackles the study of NC in quantum mechanical system using coordinate transformation~\cite{gouba}. This method has been successfully reported in~\cite{Shah} for the case of harmonic oscillator. The starting point of treating the Morse oscillator is referring to~\cite{shi,avram}.

This paper is organised sectionally where in Section 2, an introductory bit on Morse potential is presented. This section is classified into two subsections where each unfolds one-dimensional and two-dimensional Morse potential respectively. In Section 3, the development of deformed two-dimensional ladder operators is introduced.  In this section, deformed operators with respect to NC features are considered, which later lead to the alteration of the energy spectrum of the potential. The Casimir operator associated to the algebra is also reviewed. Concluding remarks are featured in the final section.

\section{The Morse oscillator}
\subsection{One-dimensional case}

Morse oscillator has been referred to as a system that describes anharmonicity in molecular interaction. Like the harmonic oscillator, Morse oscillator has its region that illustrates harmonicity but only to a certain energy level. Beyond that level are the unbound states where bond breaking starts taking place. In this work, we are interested in studying the bound states of Morse oscillator. In one-dimensional cases, Morse potential can be described as 
\begin{equation}
V(x)=V_{0}(e^{-2\alpha x}-2e^{-\alpha x}),
\end{equation}
where $V_{0}$ represents the potential depth, $\alpha$ is a constant related to the range of the potential, such that $\alpha$ is inversely proportional to the width of the potential and $x$ is the (relative) position. The total energy equation or Hamiltonian of Morse oscillator is then written as
\begin{equation}\label{Hm}
H_{M} = \frac{p^{2}_{}}{2\mu} + V(x),
\end{equation}
where $p$ is the momentum operator, $\mu$ the reduced mass, and $V(x)$ the Morse potential.
With equation (\ref{Hm}), the time-independent Schr$\ddot{o}$dinger equation is simply
\begin{equation}
-\frac{d^2}{dx^2}\phi(x) + V(x)\phi(x) = E(x)\phi(x), 
\end{equation}
where the exact solution namely $\phi(x)$ can be obtained to be \cite{bordoni}, 
\begin{equation}\label{wavefunction}
\phi^\sigma_n(x)=N^\sigma_n y^\sigma e^{-y/2}L^{2\sigma-1}_n(x), \ \ \ \  \text{for} \  n=0,1,2,...,
\end{equation}
after a change of variable of $y = ve^{- \alpha x}$, where $N^\sigma_n$ is the normalisation constant namely $N^\sigma_n = \sqrt{\frac{\alpha n!}{\Gamma (2\sigma +n)}}$, $L^{2\sigma-1}_n(x)$ is the associated Laguerre polynomial and $\sigma$ is a real parameter such that $\sigma > 0$. Equation (\ref{wavefunction}) is also called the normalized wavefunction of Morse potential. Interestingly, the significance of this wavefunction includes the derivation of ladder operators. In the context of Morse potential, ladder operators are composed of two types, namely creation operator, which is also known as a step-up operator and annihilation operator, which is a step-down operator. As the names suggest, these operators are responsible to raise or lower the energy as it goes up and down the vibrational diatomic molecular spectrum. The ladder operators are derived to have the following form
\begin{equation}\label{ladop1}
\begin{split}
K_- &= (q I + n) - \cfrac{y}{2} + \cfrac{i p_y}{\hbar \alpha}, \\
K_+ &= (q I + n) - \cfrac{y}{2} - \cfrac{i p_y}{\hbar \alpha}, \\
\end{split}
\end{equation}
where $q$ is a real quantity related to the number operator, $I$ is a unit operator and $p_y$ denotes the momentum operator. The generators or eigenvalues are calculated to be 
\begin{equation}
\begin{split}
K_- \phi_n&= C_n \phi_{n-1} , \\
K_+ \phi_n  &= C_{n+1} \phi_{n+1}, \\
\end{split}
\end{equation} 
with $K_- $ analogous to annihilation operator, $K_+$ corresponding to creation operator and $n$ as the principle quantum number   where $C_n$ takes the value of $\sqrt{n(n+2q -1)}$. This later gives rise to the associated algebra with respect to the one-dimensional system that generates $\mathfrak{su}(2)$ algebra where \cite{novaes}
\be 
[K_{-},K_{+}] = 2K_{0}, \qquad [K_{0},K_{\mp }] = \mp K_{\mp }, 
\ee
such that $K_{0}=qI$. From (\ref{ladop1}), we can see that $K_{-}$ and $K_{+}$ are in terms of $K_{0}$. It is also related to the number operator in the following form \cite{shi}

\[
K_{0} = \Big( n - \frac{\nu -1}{2}\Big),
\]

for $n\in\N$, $\nu = \sqrt{\frac{8\mu V_{0}}{\alpha^{2}\hbar^{2}}}$ and $s = \sqrt{\frac{-2 \mu E}{\alpha^{2}\hbar^{2}}}$ such that
\be
2s + 1 -\nu = -2n.
\ee

The Casimir operator of the 3-dimensional rotation group analog of the total angular momentum is given by \cite{shi}
\begin{equation}
C= K^2_0- \frac{1}{2}(K_+K_- + K_-K_+).
\end{equation}

\subsection{Two-dimensional case}

From these known results, we are instilled to further a detailed study to the case of two-dimensional space, for which the construction of Morse oscillator to higher dimensions is pretty straightforward. This is necessary since one needs at least a 2-dimensional system to apply the noncommutative quantum mechanics structure. We consider a two-dimensional Morse oscillator model obtained by superposition of two one-dimensional isotropic Morse oscillators. In 2D, we have the Morse potential to be 
\be \label{eq:1}
V(x_i)=V_{0}(e^{-2\alpha x_{i}}-2e^{-\alpha x_{i}}), 
\ee
and the Hamiltonian is in the form \cite{avram}
\be \label{eq:1}
H_{M_i} = \frac{p^{2}_{i}}{2\mu} + V(x_i), \qquad \mbox{for} \ \ i=1,2.
 \ee
Evidently, the Hamiltonian of two-dimensional Morse potential is simply the addition of two one-dimensional Morse oscillators \cite{avram}
\be \label{eq:2.2}
H_{M} = H_{M_1} + H_{M_2}.
\ee
Applying (\ref{ladop1}), we rewrite the operator algebra for the two-dimensional Morse oscillator model to be \cite{bordoni}
\begin{equation} \label{eq:2}
\begin{split}
K_{-i} &= qI - \cfrac{y_i}{2} + \cfrac{i p_{y_i}}{\hbar \alpha}, \\
K_{+i} &= qI - \cfrac{y_i}{2} - \cfrac{i p_{y_i}}{\hbar \alpha}, \\
\end{split}
\end{equation} 
for $i= 1,2.$ The corresponding algebra is thus the following  
\be \label{eq:4}
[K_{-i},K_{+i}] = 2K_{0i}, \qquad [K_{0i},K_{\mp i}] = \mp K_{\mp i}.
\ee
$K_{-i}$, $K_{+i}$ and $K_{0i}$ are ladder and number operators respectively which are written in terms of coordinate $y_i$ and momentum $p_{y_i}$ operators. We define 
\be
y_i \equiv ve^{-\alpha x_i}, \ \ \ \ p_{y_i} \equiv -i \hbar \cfrac{d}{d x_i}. \\
\ee
Now, we impose the noncommutativity feature to the commutator of the spatial coordinates, $\hat{x}_i$ by using the Baker-Campbell-Hausdorf formula to show
\be
[\hat{y}_i, \hat{y}_j ]=ve^{-\alpha(\hat{x}_i+\hat{x}_j)}(e^{\theta/2} - e^{-\theta/2} ), \ \ \ \ \ [\hat{y}_i, \hat{y}_i ] = [\hat{y}_j, \hat{y}_j ] = 0,
\ee
for $i, j = 1, 2,$ in the case of noncommutative two-dimensional quantum mechanics where $\theta$ is a non-zero noncommutative parameter. Note that from the commutative ladder operators in (\ref{eq:2}), coordinate $y_i$ and momentum $p_{y_i}$ operators can also be expressed in terms of ladder operators
\be
y_i= 2K_{0i} - (K_{+i} + K_{-i}), \ \ \ \  p_{y_i}= \cfrac{i \hbar \alpha}{2}(K_{+i} - K_{-i}),
\ee
where the commutator of the operators is simply
\be 
[y_i, p_{y_i}]= -i \hbar \alpha y_i,
\ee
given that $[p_{y_i}, p_{y_j}]=0.$ The Hamiltonian of equation (\ref{eq:1}) can then be rewritten as \cite{molnar}
\be \label{eq:3}
{H}_{M_i}= \cfrac{\hbar^2 \alpha^2}{2 \mu} (K_{+i}K_{-i}- K_{0i}^2).
\ee
With (\ref{eq:3}), we can then compute the commutator of ${H}_{M_i}$ with $K_{-i}$ and $K_{+i}$ as follows
\be
\begin{split}
[{H}_{M_i}, K_{-i}] &= - \cfrac{\hbar^2 \alpha^2}{2 \mu} (3K_{0i}K_{-i} + K_{-i}K_{0i}), \\
[{H}_{M_i}, K_{+i}] &= \cfrac{\hbar^2 \alpha^2}{2 \mu} (3K_{+i}K_{0i} + K_{0i}K_{+i}).
\end{split}
\ee  
Taking $H_{M_i}\phi_{n,m}=E_{n,m}\phi_{n,m}$, one can then easily find the energy eigenstates. From the relations above, the eigenstate $K_{-i} \phi_{n,m}$ shows the action of lowering the energy $E_{n,m}$ while the eigenstate $K_{+i} \phi_{n,m}$ exhibits the action of creating or raising $E_{n,m}$. These results are consistent with the role of ladder operators appear in a system with oscillatory motion. Comparatively, one may write the two-dimensional Casimir operator as
\begin{equation}
C_i= K^2_{0i}- \frac{1}{2}(K_{+i}K_{-i} + K_{-i}K_{+i}), \ \ \ \ \  \text{for} \ \ i= 1,2.
\end{equation}

\section{Deformed Morse Operators}

In this section, we will unravel the idea of deformed ladder operators. It is well-known that ladder operator method is the most convenient tool to solve exactly solvable systems for obtaining the stationary states from the time-independent Schr$\ddot{o}$dinger equation. For the deformed ladder operators arise in the NC space, results show that they can be written as a linear combination of the ordinary ladder operators of the commutative system~\cite{Shah}. For the case of deformed operators, each state constructed describes the associated discrete state representing diatomic molecular vibration in the NC configuration space. To begin with,  deformed Morse operators are parameterised by a noncommutative parameter which is a $n \times n$ matrix for any $n$-dimensional case. It is worth to note that the lowest dimension would be 2D in order to verify the noncommutativity effect. Similar to the case of the commutative system, the deformed creation operators will generate the generalised eigenfunctions, namely the states for the NC Morse oscillator.  For the two-dimensional case, we fix one of the two distinctive states to be unvaried while the other being operated with the ladder operator.  The action of the deformed operator is the addition of ordinary ladder operator imposed on each of the two states individually. This translates to a non-simultaneous operation of the deformed ladder operators on the states. This can be described mathematically by
\begin{equation} \label{eq:5}
\begin{split}
{K}^g_{-1} \phi_{n,m} &= g_{11} K_{-1}  \phi_{n-1,m} + g_{12} K_{-2}  \phi_{n,m-1},  \\
{K}^g_{-2} \phi_{n,m} &= g_{21} K_{-1}  \phi_{n-1,m} + g_{22} K_{-2}  \phi_{n,m-1},  \\
{K}^g_{+1} \phi_{n,m} &= g_{11} K_{+1}  \phi_{n+1,m} + g_{12} K_{+2}  \phi_{n,m+1},  \\
{K}^g_{+2} \phi_{n,m} &= g_{21} K_{+1}  \phi_{n+1,m} + g_{22} K_{+2}  \phi_{n,m+1},  \\
\end{split}
\end{equation}
with $n,m$ are integers and the deformed ladder operators are in terms of the ordinary ladder operators having being operated with some matrix $g$ with elements $g_{ij} \in$ $GL(2,\mathbb{C})$ such that \cite{Shah}
\begin{equation}
g = 
\begin{pmatrix}
    g_{11} & g_{12}\\
    g_{21} & g_{22} \\
\end{pmatrix}.
\end{equation} 
Matrix $g$ with specific form will give an explicit example of NCQM in problems as discussed in \cite{Shah}. Note that for the standard way to achive noncommutativity is by particular coordinate transformation~\cite{gouba} of the commutative coordinates but in this case, we apply the transformation involving the (ladders) operators. Thus for
\begin{equation} \label{eq:7}
g = 
\begin{pmatrix}
   1 & 0 \\
   0 & 1 \\
\end{pmatrix},
\end{equation} 
the case of NCQM becomes the case of ordinary QM such that
\begin{equation} \label{eq:6}
\begin{split}
{K}^g_{-1} \phi_{n,m} &=  K_{-1}  \phi_{n-1,m} ,  \\
{K}^g_{-2} \phi_{n,m} &=  K_{-2}  \phi_{n,m-1},  \\
{K}^g_{+1} \phi_{n,m} &= K_{+1}  \phi_{n+1,m} ,  \\
{K}^g_{+2} \phi_{n,m} &= K_{+2}  \phi_{n,m+1},  \\
\end{split}
\end{equation}
for two-dimensional problems. Note that the usual transformation of NC coordinate can be found in \cite{bertolani}.

The deformed form of these operators is indicated by $g$  to differentiate it from the ordinary form. The corresponding two-dimensional deformed generators are thus
\begin{equation}
\begin{split}
{K}_{-1}^g \phi_{n,m} &= g_{11} C_n \phi_{n-1,m} + g_{12} C_m \phi_{n,m-1},  \\
{K}_{-2}^g \phi_{n,m} &= g_{21}C_n  \phi_{n-1,m} + g_{22} C_m  \phi_{n,m-1},  \\
{K}_{+1}^g \phi_{n,m} &= g_{11} C_{n+1}\phi_{n+1,m} + g_{12} C_{m+1} \phi_{n,m+1},  \\
{K}_{+2}^g \phi_{n,m} &= g_{21} C_{n+1} \phi_{n+1,m} + g_{22} C_{m+1}\phi_{n,m+1},  \\
\end{split}
\end{equation} 
for $C_{n}= \sqrt{n(n+2q -1)}$  and  $C_{m}= \sqrt{m(m+2q -1)}$ which correspond to the energy spectra of the potential. 
Here, (\ref{eq:5}) shows a linear combination of ordinary ladder operators $K_{\pm}i$. It is also worth to mention that the deformed number operator generates the ordinary number operator such that
\be
{K}^g_{0i}\phi^{\sigma}_{n,m}={K}_{0i}\phi^{\sigma}_{n,m}.
\ee
From the operators constructed and (\ref{eq:4}), we can compute the commutation relations of the deformed ladder operators which are in terms of the ordinary form.  The only survived commutation relations are
\begin{equation}
\begin{split}
[{K}^g_{\mp i},{K}^g_{\pm i}] &= \pm 2[g^{2}_{ii}K_{0i}+g^{2}_{ij}\,K_{0j}] , \\
[{K}^g_{\mp i},{K}^g_{\pm j}] &= \pm 2[g_{ii}g_{ji}K_{0i}+g_{ij}g_{jj}\,K_{0j}] , \\
[{K}_{0i},{K}^g_{\mp i}] &= \mp g_{ii}K_{\mp i},\qquad [{K}_{0i},{K}^g_{\mp j}] = \mp g_{ij}K_{\mp i}. \\
\end{split}
\end{equation}
The deformed Casimir operator is thus
\begin{equation}
C_i^g= K^2_{0i}- \frac{1}{2}(K^g_{+i}K^g_{-i} + K^g_{-i}K^g_{+i}).
\end{equation}
The resulting commutation relations can be shown as follows
\be \label{defCas}
\begin{split}
 [C_i^g, K_{0i}] = & \frac{1}{2}g_{ii}g_{ij}[-K_{-i}K_{+j}+K_{+i}K_{-j}\\
 &+K_{-j}K_{+i}-K_{+j}K_{-i}], \\
  [C_i^g, K_{0j}] = & \frac{1}{2}g_{ii}g_{ji}[+K_{-i}K_{+j}-K_{+i}K_{-j}\\ &-K_{-j}K_{+i}+K_{+j}K_{-i}], \\
[C_i^g, K^g_{\mp i}]= & \mp g_{ii}[K_{0i}K_{\mp i} + K_{\mp i}K_{0i}] \\ & \pm [g^2_{ii}K_{0i} + g^2_{ij}K_{0j}] (g_{ii}K_{\mp i} + g_{ij}K_{\mp j})  \\
& \pm (g_{ii}K_{\mp i} + g_{ij}K_{\mp j}) [g^2_{ii}K_{0i} + g^2_{ij}K_{0j}] ,\\
 [C_i^g, K^g_{\mp j}]= & \mp g_{ij}[K_{0i}K_{\mp i} + K_{\mp i}K_{0i}] \\ & \pm [g_{ii}g_{ji} K_{0i} + g_{ij}g_{jj}K_{0j}] (g_{ii}K_{\mp i} + g_{ij}K_{\mp j}) \\
 & \pm (g_{ii}K_{\mp i} + g_{ij}K_{\mp j})[g_{ii}g_{ji} K_{0i} + g_{ij}g_{jj}K_{0j}].\\
\end{split}
\ee

Unlike the ordinary Casimir operator which carries the element of the centre of the universal enveloping algebra that commutes with the rest of the element of the algebra, the deformed Casimir parameterised by $g$ is not particularly the case. It is shown that, since $C_{i}^{g}$ is not Casimir invariant, it is just another element of the whole algebra of the deformed operator that makes up $\mathfrak{su}(2)$. \\

In addition to that, we can now write the deformed Hamiltonian

\begin{equation}
\begin{split}
{H}^g_{M} &= {H}^g_{M_1} + {H}^g_{M_2},\\
 &= \cfrac{\hbar^2 \alpha^2}{2 \mu} [{K}^g_{+1}{K}^g_{-1} - ({K}^g)^2_{01}] + \cfrac{\hbar^2 \alpha^2}{2 \mu} [{K}^g_{+2}{K}^g_{-2} - ({K}^g)^2_{02}], \\
&= \cfrac{\hbar^2 \alpha^2}{2 \mu} [(g_{11} K_{+1}  + g_{12} K_{+2} )(g_{11} K_{-1}  + g_{12} K_{-2} ) - {K}^2_{01}] \\
& + \cfrac{\hbar^2 \alpha^2}{2 \mu} [(g_{21} K_{+1}  + g_{22} K_{+2} )(g_{21} K_{-1}  + g_{22} K_{-2} ) - {K}^2_{02}],\\
&= \cfrac{\hbar^2 \alpha^2}{2 \mu} [(g^2_{11}+g^2_{21}) K_{+1} K_{-1} + (g_{11}g_{12} + g_{12}g_{22})\\ & \times ( K_{+1} K_{-2} + K_{+2} K_{-1}) + (g^2_{12}+g^2_{22} ) K_{+2} K_{-2}\\
& - {K}^2_{01}-{K}^2_{02}] ,
\end{split}
\end{equation}
where noncommutative quantum mechanics in two dimensions is considered. The deformed Hamiltonian can be transformed to the ordinary Hamiltonian, taking (\ref{eq:7}) so it becomes
\begin{equation}
\begin{split}
{H}_{M} &= \cfrac{\hbar^2 \alpha^2}{2 \mu} [ K_{+1} K_{-1} - {K}^2_{01}] + \cfrac{\hbar^2 \alpha^2}{2 \mu} [K_{+2} K_{-2} - {K}^2_{02}].
\end{split}
\end{equation}
Here, we show that by operating a specific matrix $g$, that is, a unit operator, the deformed Hamiltonian is converted back to its ordinary form as in (\ref{eq:2.2}). \\

\section{Conclusion}

In the present work, we have analysed the algebra of the deformed ladder operators for the case of noncommutative Morse oscillator. These generalised operators are important as they are a tool to generate the allowed bound states for the NC Morse oscillator. The method shown here is different from the common method to transform the commutative configuration space to the noncommutative configuration space. Note that the generalised algebra appear in (\ref{eq:5}) which is parametrised by matrix $g_{ij}$ gives a new type of deformed Casimir operator of the algebra which is non-invariant (\ref{defCas}). From the computation we could arrive to our deformed Hamiltonian in two dimensions which directly link to the originally ordinary ladder operators of the potential. We also prove that for a specific $2$ by $2$ matrix, the Hamiltonian deduce to the ordinary Morse oscillator in 2D. Further work on NCQM is in the draft where we tackle the problem from a different angle, which is by utilisation of polynomial which describes the states of NC Morse oscillator. \\

\section*{Acknowledgements}
We would like to acknowledge and thank UPM for supporting this research through UPM-IPM Grant (9473100) and GP-IPS Grant (9645700).

\end{document}